\documentclass[sigconf,nonacm]{acmart}
\usepackage{lscape} 
\AtBeginDocument{%
  \providecommand\BibTeX{{%
    \normalfont B\kern-0.5em{\scshape i\kern-0.25em b}\kern-0.8em\TeX}}}

\copyrightyear{2025}
\acmYear{2025}
\setcopyright{rightsretained}
\acmDOI{}
\acmISBN{}

\begin{document}

\title{Ephemeral Feeds and Enduring Rituals: RushTok and the Formation of Event-Based Algorithmic Communities}

\author{Emelia May Hughes}
\email{ehughes8@nd.edu}
\affiliation{%
  \institution{University of Notre Dame}
  \city{Notre Dame}
  \state{Indiana}
  \country{USA}
}

\author{Tim Weninger}
\email{tweninger@nd.edu}
\affiliation{%
  \institution{University of Notre Dame}
  \city{Notre Dame}
  \state{Indiana}
  \country{USA}
  }

\begin{abstract}
Each August, TikTok’s For You page turns the University of Alabama’s sorority recruitment into RushTok. We examine RushTok as an event-based algorithmic community: a collective assembled around a bounded offline ritual and sustained by recommendation. Using a mixed-methods survey (n=71) and a reflexive account of creator outreach, we ask who participates, how, and with what stakes. Findings show an ambiguous and entertainment based throughline; many called it a community (51/71) but few claimed membership (11/71). Affiliation centered on creators rather than shared practices, with parasocial attention clustering around a small set of potential new members (PNMs) and returning figures. Higher content exposure tracked with self-identification as a community member; those members commented, followed creators, and engaged across videos. Attempts to interview creators were met with silence or refusals, reflecting community boundary-work despite viral visibility. We outline implications for platform governance, including time-bounded context, graduated visibility, and aftercare.
\end{abstract}

\maketitle

\section{Introduction}

Short-form video platforms such as TikTok have rapidly become key infrastructures for cultural exchange, news circulation, and community formation online. In contrast to earlier social platforms built around explicit friend networks or group affiliations, TikTok organizes participation through algorithmic curation, most notably its \textit{For You} page \cite{Cheney-Lippold-NewAlgorithmicIdentity-2011,EslamiEtAl-FirstItThen-2016,DeVitoEtAl-AlgorithmsRuinEverything-2017,DeVitoEtAl-HowPeopleForm-2018,DeVito-AdaptiveFolkTheorization-2021}. Rather than joining predetermined groups, users encounter content and people through algorithmically shaped pathways, giving rise to large, transient collectives whose boundaries and norms emerge in practice. These collectives increasingly influence offline life, from coordinated online actions that affected attendance at a high-profile political rally \cite{BandyDiakopoulos-TulsaFlopCaseStudy-2020} to the market effects associated with BookTok \cite{Harris-HowTikTokBecame-2022}, elevating otherwise local or niche rituals into widely shared spectacles. Yet the processes by which these temporary formations arise, sustain attention, and connect to offline identities remain poorly understood.

This paper focuses on \textit{RushTok}, the algorithmically assembled collective that forms around the University of Alabama's sorority recruitment. Each August, thousands of TikTok users become entralled by ''OOTD'' (outfit of the day) videos, behind-the-scenes glimpses of recruitment rituals, and meta-commentary that reframes a campus tradition as a shared digital event. Unlike traditional communities of practice \cite{Wenger-CommunitiesPracticeLearning-1999} or identity-based groups such as fandoms \cite{Baym-TuneLog-2025}, RushTok illustrates what we term an \textit{event-based algorithmic community}: a formation organized by spectacle, temporality, and algorithmic amplification rather than durable ties \cite{CollieWilson-Barnao-PlayingTikTokAlgorithmic-2020}. RushTok shows how people can feel part of a community without ever explicitly choosing to join it. Drawing on a mixed-methods study that combines a survey of RushTok viewers with an auto-ethnographic reflection on recruiting creators for interviews, we examine how participants experience, narrate, and distance themselves from RushTok, and how the collective's insularity both sustains its boundaries and resists outside scrutiny. In doing so, we build on theories of community and mediated collectives in HCI \cite{Papacharisi-NetworkedSelfIdentity-,Turner-RitualProcessStructure-2011} and identify implications for how platforms surface and govern such emergent formations.

\begin{figure}
\centering
\includegraphics[width = 2.5in]{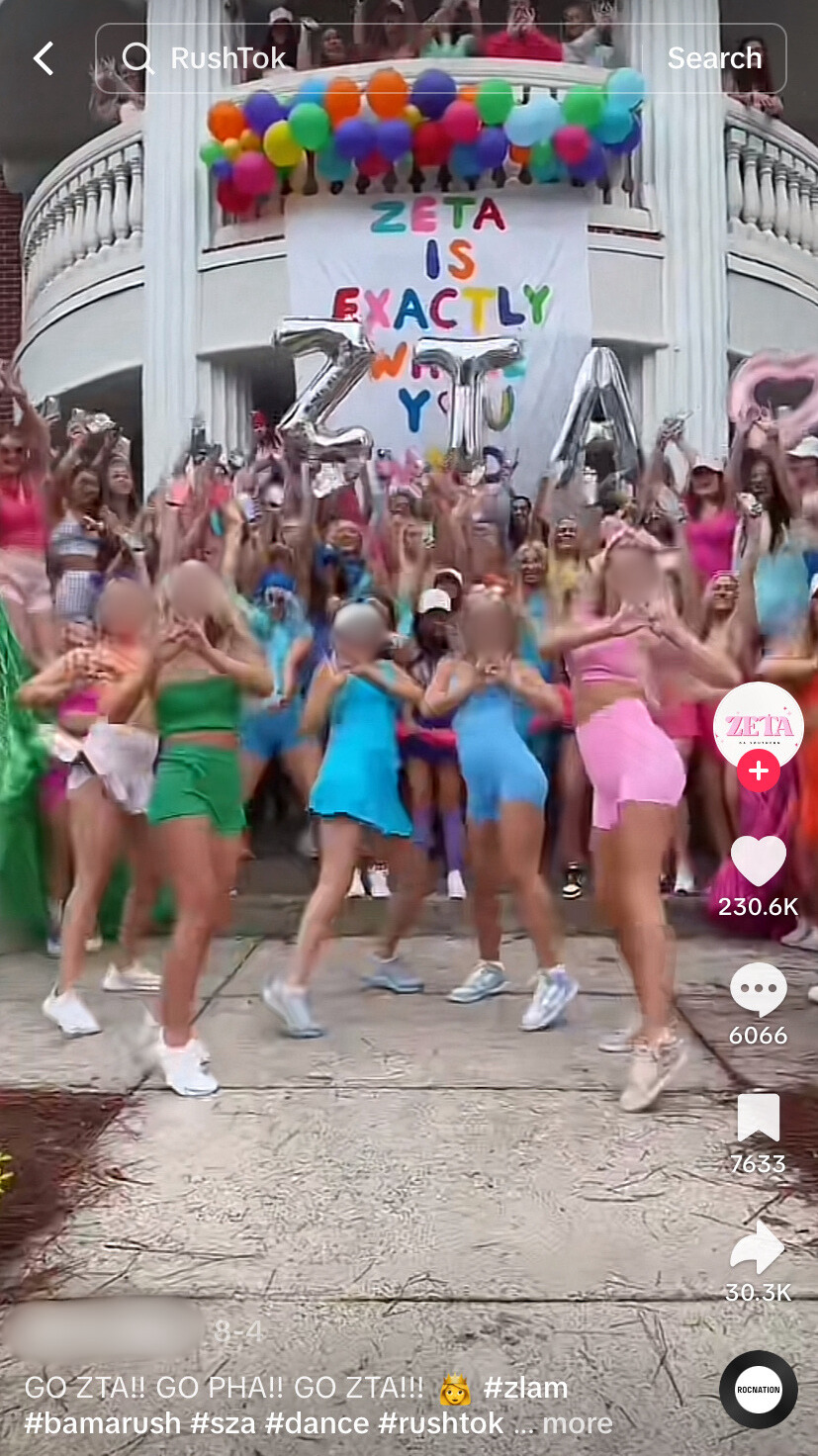}
\caption{Members of a University of Alabama sorority participate in a \#RushTok dance trend with coordinated choreography and bright outfits; faces are blurred.}
\label{fig:rushtok}
\end{figure}

\paragraph{Context: Southern sorority recruitment}
Sorority recruitment at large Southern universities unfolds as a structured, multi-round ritual in which potential new members (PNMs) rotate through sorority houses, engage in highly choreographed conversations, and present carefully curated appearances. Each local sorority house, called a Chapter, manages a public image that blends tradition, philanthropic messaging, and aesthetic performance. Reputational hierarchies and alumni expectations shape decisions, and selection culminates in preference rounds and bid day. The process concentrates attention, status, and risk, making it an especially visible and consequential campus event. Against this backdrop, TikTok's visibility infrastructures amplify not only style and performance but also evaluation and surveillance, turning a localized rite into a mass-mediated spectacle.

Much of what we know about online communities comes from networked, fandom, or identity-based models in which belonging is explicit, opt-in, and sustained. By contrast, TikTok collectives are assembled through algorithmic circulation: users do not consciously join so much as get pulled in by the \textit{For You} page. These event-based algorithmic communities pose several questions: (1) how belonging is experienced when membership is ambiguous; (2) what motivates participation when exposure is largely algorithmic; and (3) how such formations reshape the offline rituals they orbit. Conventional methods such as interviews, ethnography, and network analysis map unevenly onto short-lived or insulated collectives, leaving important gaps in understanding how visibility and boundaries are produced and governed \cite{CollieWilson-Barnao-PlayingTikTokAlgorithmic-2020}.

We address this gap through a case study of RushTok, the TikTok surge that forms each August around the University of Alabama's sorority recruitment. RushTok transforms a localized campus ritual into a global spectacle via algorithmic amplification, providing a concrete case for examining how event-based algorithmic communities operate in practice. We surveyed viewers to capture engagement patterns, self-identification, and motivations; open-text reflections ranged from playful spectatorship to parasocial attachment, belonging, ambivalence, and distance. In parallel, efforts to interview RushTok creators often met with silence or refusals, which we interpret as boundary-work and selectivity. Taken together, these perspectives allow us to characterize RushTok as an event-based algorithmic community and to highlight broader implications for how algorithms assemble audiences around offline events, how ambiguous participation is negotiated, and how boundary-work shapes the experience of both participants and researchers.

\paragraph{Findings in Brief} RushTok complicates what it means to belong online. Many respondents described it as a community, yet far fewer claimed membership, underscoring the tension between spectatorship and belonging when algorithmic circulation drives exposure. Most engagement was passive and entertainment-oriented. Situated ties were salient: PNMs, alumni, and parents referenced sorority connections, while others expressed nostalgia, curiosity, or critique. Ambivalence and contestation, as much as belonging, sustained participation, distinguishing RushTok from networked or identity-based communities organized around reciprocal ties. TikTok visibility also fed back into the offline ritual, amplifying anticipation, reputational stakes, and pressure, thereby raising the perceived stakes of recruitment. Framed as an event-based algorithmic community, RushTok offers a tractable case for how algorithms assemble audiences, how ambiguous participation is negotiated, and how boundary-work constrains both participants and researchers. These dynamics carry design and governance implications: algorithmic surges can mint instant influencers, reshape the stakes of physical events, and expose participants to risks without clear institutional safeguards.

\section{Related Work}
We synthesize four strands: (1) algorithmic identity and folk theories, (2) TikTok's memetic logics, (3) communities and algorithmic resistance, and (4) event-based temporality. Together these works motivate our focus on \textit{event-based algorithmic communities}. Prior work explains how individuals infer, negotiate, and resist algorithmic systems. What remains under-specified is how these dynamics scale from individual tactics to temporally bounded collectives that crystallize around offline rituals.

\subsection{Algorithmic Identity and Folk Theories}

Work on \textit{algorithmic identity} shows how platforms infer and modulate who users are through datafied traces \cite{Cheney-Lippold-NewAlgorithmicIdentity-2011}. Identity operates as a moving target of classification, where categories are computed, updated, and leveraged for relevance, recommendation, and moderation. Complementary research on \textit{folk theories} documents how people narrate and reason about opaque systems, forming working explanations that guide everyday tactics such as what to post, when to watch, and when to like or refrain \cite{GelmanLegare-ConceptsFolkTheories-2011, ToffNielsen-JustGoogleIt-2018, RaderGray-UnderstandingUserBeliefs-2015, TainaBucherBucher-AlgorithmicImaginaryExploring-2017, Seaver-AlgorithmsCultureTactics-2017}. These explanations function as {behavior-shaping heuristics}: people test hypotheses about what the system privileges, adjust when results deviate, and refine tactics as platforms change \cite{DeVitoEtAl-AlgorithmsRuinEverything-2017, DeVitoEtAl-HowPeopleForm-2018, EslamiEtAl-FirstItThen-2016, EslamiEtAl-AlwaysAssumedThat-2015, EslamiEtAl-CommunicatingAlgorithmicProcess-2018, DeVito-AdaptiveFolkTheorization-2021, KarizatEtAl-AlgorithmicFolkTheories-2021, SimpsonSemaan-YouYouEveryday-2021}. Personalization and inferred categorization further structure how people come to see themselves as targets of algorithmic classification across platforms \cite{BoekerUrman-EmpiricalInvestigationPersonalization-2022, Wang-CalculatingDatingGoals-2020, ZengEtAl-ResearchPerspectivesTikTok-2021}.

Crucially, folk theories are \textit{social objects} as well as individual mental models. They travel in comment sections, creator monologues, and tutorial-style meta-videos about how to reach the \textit{For You} page, and they are negotiated in replies and stitches that evaluate whether a tactic works. Explanations of the algorithm circulate alongside the very content they aim to optimize. In the context of RushTok, we attend to how these interpretive tactics operate within a compressed event window and become visible as shared talk, recurring practices, and recognizable cues of belonging. During such windows, folk theories can coordinate participation at public scale, not through formal rules, but through widely legible heuristics about exposure, timing, and trend alignment.

\subsection{TikTok's Memetic Logics}

TikTok's platform grammars privilege {replication and templating}, including sounds, duets, stitches, and visual formats, organizing what scholars call \textit{imitation publics} \cite{ZulliZulli-ExtendingInternetMeme-2020, Conte-MemesSocialMinds-2001}. Trend participation is scaffolded by ready-made structures: a sound encodes timing, a duet frame encodes position, and a stitch encodes call-and-response. Accounts of the \textit{algorithmized self} argue that identity work is routed through recommendation logics, such that performers shape contributions for audiences and for the system that assembles those audiences \cite{BhandariBimo-TIKTOKALGORITHMIZEDSELF-2020}. Media and platform studies emphasize \textit{algorithmic culture} and \textit{programmability} as conditions that steer creative contributions and social interaction, where production choices are entangled with expectations about ranking, distribution, and the costs of being ignored \cite{CollieWilson-Barnao-PlayingTikTokAlgorithmic-2020, DijckPoell-UnderstandingSocialMedia-2013}. Reporting on TikTok's growth underscores the salience of {real-time interest signals} for circulation \cite{Byford-HowCompanyTikTok-2018, Harris-HowTikTokBecame-2022}. These dynamics sit within a longer lineage of memetic diffusion and cultural replication (cf. \cite{Hawkins-MemeMachine-2000}). Platform policies and public communications also shape these dynamics, including transparency claims and governance controversies \cite{-CommunityGuidelinesTikTok-2023, Pappas-TikTokLaunchTransparency-2019, Perez-TikTokOpenTransparency-2020, Doffman-WarningAppleSuddenly-, Reuters-TikTokChinaByteDance-2020, Hansler-UnitedStatesLooking-2020, Schuman-WhyAmericaAfraid-2020}.

Memetic grammars matter for {temporal coordination}. Shared audio, countdown motifs, and serial formats such as day-by-day updates and OOTD cycles give dispersed creators a common clock and give viewers a common reading frame. The recommendation pipeline can then align attention around these grammars. As interest signals accrue, similar videos cluster, and a viewing public coalesces, followed by rapid drift as the system moves on. In RushTok, templating provides the scaffolding through which the recommender synchronizes attention with the offline cadence of recruitment. The result is a short-lived, synchronized spectacle, a surge of coordinated watching, shared evaluation criteria, and quickly diffusing in-jokes that map onto the week-long temporal arc of the event. This synchronization is consequential in Southern sorority recruitment, where chapters curate aesthetic and reputational images under alumni oversight and perceived hierarchies, and where presentation, choreography, and tradition carry social stakes that the platform can magnify.

\subsection{Communities and Algorithmic Resistance}

A substantial body of work documents how communities mobilize {against} algorithmic impositions. Studies of adolescents' interactions with short-video recommendation systems surface avoidance, selective engagement, and tactical obfuscation \cite{LvEtAl-AdolescentsAlgorithmicResistance-2022}. Conceptual work elaborates \textit{productive resistance} and \textit{repair} as responses to constraint and bias \cite{VelkovaKaun-AlgorithmicResistanceMedia-2021, DosonoSemaan-DecolonizingTacticsCollective-2020}. Media scrutiny of moderation controversies shows how uneven enforcement prompts experimentation with workarounds and calls for transparency \cite{Botella-TikTokAdmitsIt-2019, Dias-TikTokToldModerators-2020, Robertson-TikTokPreventedDisabled-2019}. Work on identity-based and activist collectives shows how hashtags sustain visibility and action, establishing well-studied paths to cohesion via grievance \cite{LeeLee-StopAsianHateTikTokAsian-2023, RayEtAl-FergusonDeathMichael-2017, BrownEtAl-SayHerNameCaseStudy-2017, BandyDiakopoulos-TulsaFlopCaseStudy-2020}. Adjacent literatures examine professional and interest-based collectivities that cultivate belonging and knowledge exchange within platform constraints \cite{LiuEtAl-SelfiesSocialMovements-2017, Vizcaino-VerduAbidin-TeachTokTeachersTikTok-2023, Petrovic-KaraokeLipsyncingPerformance-2023, HiebertKortes-Miller-FindingHomeOnline-2023, BocciaArtieriEtAl-RadarPrivateGroups-2021}. These tensions are framed within broader accounts of platform politics and the trade-offs of compelling design \cite{Tarleton-PoliticsPlatforms-2010, DeLosSantosKlug-TikTokTradeoffCompelling-2022}, as well as the role of humor and affect in civic practice \cite{SchaadhardtEtAl-LaughingDontCry-2023}.

While resistance-based collectives are well documented, RushTok exemplifies a parallel formation that is not anchored in grievance. Here cohesion arises from anticipation, evaluation, in-jokes, and \textit{parasocial spectatorship}. Participation often takes the form of watching together rather than formal membership. Contribution can be lightweight, for example liking, commenting, or stitching commentary, yet still consequential for visibility. Such collectives also contain \textit{outsider positions}, where participants insist they are just watching or not part of this, even as their engagement sustains the spectacle. Boundary work operates in both directions. Insiders articulate norms about appropriate presentation and how to read signals. Outsiders articulate distance and critique, all within the same algorithmically assembled crowd. These dynamics complicate researcher access. When collectives are convened through spectacle rather than durable ties, contact pathways are shallow and gatekeeping is diffuse, a point we return to in our auto-ethnographic reflection. For broader syntheses of platform-wide patterns, see \cite{ZengEtAl-ResearchPerspectivesTikTok-2021}.

\subsection{Event-Based Communities and Temporality}
Sorority rush is a short, scheduled ritual. Classic work on ritual treats recruitment as a liminal passage in which roles are tried on and futures are negotiated. Media-events scholarship shows how mediation creates shared attention. Broadcast and platform formats pull dispersed spectators into a common present, give ordinary acts symbolic weight, and link remote audiences to the stakes of an unfolding rite \cite{CouldryEtAl-MediaEventsGlobal-2009, -MediaEvents-}. Platform research on networked identity explains how collectives form around shared grammars and values \cite{Papacharisi-NetworkedSelfIdentity-}.

TikTok collectives differ from durable communities built through membership and obligation. They are assembled by ranking systems and by reusable templates such as sounds, stitches, and serial formats. What holds them together is timing and repetition rather than applications or roles. Fandom and parasocial literatures help explain this pull. Viewers can feel connected to creators through one-sided intimacy and through content that is easy to share and remix \cite{HortonRichardWohl-MassCommunicationParaSocial-1956, WangShang-HowSocialParasocial-2024}. Research on ephemerality shows how time limits and scarcity shape participation and meaning-making \cite{BayerEtAl-SharingSmallMoments-2016, McRobertsEtAl-ShareFirstLater-2017, McRobertsEtAl-ItViewersAudience-2016, RobertsonEtAl-PopupsEphemeralityConsumer-2018}.

These ideas clarify RushTok. Templates and recommendation bring attention together at the same pace as the offline schedule. The ritual provides stakes and a clear arc. Parasocial viewing and easy replication bind spectators who share little else. Algorithmic amplification raises scrutiny, increases perceived pressure, and briefly makes local performers visible at scale. We use this temporal lens to interpret how RushTok forms, peaks, and recedes \cite{CouldryEtAl-MediaEventsGlobal-2009, -MediaEvents-, Papacharisi-NetworkedSelfIdentity-, HortonRichardWohl-MassCommunicationParaSocial-1956, WangShang-HowSocialParasocial-2024, BayerEtAl-SharingSmallMoments-2016, McRobertsEtAl-ShareFirstLater-2017, McRobertsEtAl-ItViewersAudience-2016, RobertsonEtAl-PopupsEphemeralityConsumer-2018}.

\section{Methodology}
\subsection{Survey Study}

\subsubsection{Participants}
We recruited $n=71$ survey respondents. Most identified as female (67, 94\%), with 3 male and 1 preferring not to disclose. Ages spanned 18 to 65+: 20 were 18--22, 21 were 23--34, 19 were 35--44, 8 were 45--54, 2 were 55--64, and 1 was 65+. Most reported no formal affiliation with the University of Alabama (UA; 63). The remainder included current UA students (3), a UA alumna/us (1), family members of sorority participants (2), and two self-described spectators (``lifelong diehard Bama fan'' and ``major football rival''). Educational attainment was varied: 26 held a bachelor's degree, 14 a master's, 4 a doctorate, and 3 a professional degree; 17 were current undergraduates, 6 had completed high school or a GED, and 1 held an associate’s degree.

\subsubsection{Recruitment}
We used two channels: algorithmic outreach on TikTok and place-based outreach near campus, mirroring RushTok's hybrid digital--physical presence. We produced a short video in the visual idiom of RushTok (outfit-of-the-day framing, text overlays, casual yet polished aesthetic) and used TikTok's \textit{Promote} feature to target audiences engaging with \#BamaRush and \#RushTok, so that the recruitment artifact traveled through the same algorithmic pathways as RushTok content. To complement this, we distributed physical business cards with survey QR codes at establishments near the UA campus, including coffee shops, bookstores, and boutiques. This two-channel approach increased ecological validity by aligning recruitment with how RushTok circulates online and how it is discussed in local spaces.

\subsubsection{Ethics and Transparency}
This study was approved by our Institutional Review Board ([\textit{redacted}] IRB; protocol [\textit{redacted}]). Participants provided informed consent before participating and could withdraw at any time without penalty. We ran three micro-budget, {paid in-platform promotion} campaigns using TikTok's \textit{Promote} tool and logged all platform-reported metrics. Two \textit{More website visits} campaigns (USD \$16 each) delivered 17{,}040 and 28{,}825 views, yielding 296 and 297 link clicks (CTR 1.7\% and 1.0\%; CPC \(\approx\)\ \$0.05). A \textit{More video views} campaign (USD \$10.89) exceeded the 1{,}600-view guarantee with 28{,}109 views (\(\sim 17.6\times\) the minimum; CPV \$0.00039). Across campaigns, TikTok reported predominantly female reach (57--77\%) and concentration among 18--34-year-olds (60--71\%), with smaller 35--44 and 13--17 segments. Targeting used the hashtags \#BamaRush and \#RushTok; no demographic exclusions beyond age 18+ were applied. In total, the video accrued 96{,}941 views during data collection (August 8, 2024 to September 4, 2024).

\subsubsection{Survey Design}
The survey prioritized low-friction participation, reflecting that many people encountered RushTok passively via the \textit{For You} page. It included:
\begin{itemize}
  \item Closed-ended items capturing demographics, general TikTok use, and frequency of exposure to RushTok.
  \item Optional open-ended prompts inviting participants to describe their experiences of RushTok, belonging, and community, recognizing that many participants were low-involvement scrollers.
\end{itemize}

\subsection{Analysis}

\subsubsection{Quantitative}
We summarized closed-ended items descriptively, including distributions of demographics, RushTok discovery channels, content types viewed, creator preferences, community self-identification, and perceived offline relevance. We then contrasted respondents who self-identified as RushTok community members ($n{=}11$) with those who did not ($n{=}60$), focusing on exposure frequency, motivations, and engagement behaviors. Analyses were limited to descriptive contrasts appropriate to our exploratory aims and modest sample size. We report counts and percentages, noting item-level denominators when they differ from $n{=}71$, and, where useful, present side-by-side distributions for the two groups. We did not conduct hypothesis tests or model-based inference; observed differences are interpreted as descriptive patterns rather than generalizable effects.

\subsubsection{Qualitative}
We conducted thematic analysis of five open-text survey questions. Three researchers independently coded all responses using a shared, iteratively refined codebook organized into six domains: {Community Atmosphere \& Culture}, {Content}, {Identity \& Belonging}, {Linkage to the Physical Event}, {Systemic/Structural Commentary}, and {Trajectory of the Community}. Responses could receive multiple codes within each domain. Codebook development followed grounded, iterative comparison and memoing. After independent coding, we produced a consensus dataset using majority rule, retaining a code when at least two of three coders applied it. Additive and base-code dependencies were enforced; rare singletons were retained only with documented rationale. To foreground transparency rather than treat agreement as validity, we summarized coder overlap in an agreement table that reported, for each code, how many coders applied it; discrepancies were then reviewed and resolved during adjudication (Appendix Table~\ref{tab:icr}). Building from the consensus set, we conducted reflexive thematic analysis to articulate cross-cutting themes (for example, entertainment as baseline, contested atmosphere, belonging versus outsider positioning, parasocial anchors, event temporality, limited offline linkage). Theme coverage is reported descriptively in the Results, and code counts from the consensus table appear in the Appendix.

\subsubsection{Integration}
We used a convergent mixed-methods design in which descriptive quantitative patterns and qualitative themes were treated as complementary evidence. Quantitative summaries (for example, recognition of RushTok as a community versus self-identification, exposure to common content types) guided targeted readings of open-text responses. In turn, emergent themes (for example, outsider positioning, parasociality) motivated additional descriptive tabulations. To make these linkages explicit, the Results include a joint display that aligns quantitative patterns, qualitative themes, and exemplar quotes, alongside a theme-summary table that lists each theme, representative codes, and its descriptive coverage across participants

\subsection{Interview Recruitment Attempts}
In parallel with the survey, we sought to recruit RushTok creators and influencers for semi-structured interviews. We identified public TikTok accounts of sorority members and potential new members and sent individualized invitations from the first author's researcher-identified account via TikTok direct messages. Where creators listed an Instagram handle or email address, we sent the same invitation through those channels; when a management contact was listed, we contacted that representative as well. The outreach script introduced the researcher, described the study purpose, emphasized confidentiality, and invited participation. In total, we sent approximately 90 messages across platforms.

Recruitment also included place-based outreach. During recruitment week in fall 2024, the first author visited the University of Alabama, distributed cards with a survey link and contact information at campus and nearby businesses, and initiated conversations with the Office of Sorority and Fraternity Life. We contacted the Panhellenic Council, which referred us to the National Panhellenic Council. We submitted materials to that body and followed up multiple times, but did not receive a response. These efforts yielded no completed interviews during the study period, a result we interpret as part of the community's boundary-work and selectivity.

\subsection{Limitations}
Our survey sample skews female and unaffiliated with UA, reflecting RushTok's audience rather than campus participants; generalizability is therefore limited to TikTok users exposed to RushTok. Recruitment relied on paid in-platform promotion and place-based convenience sampling, so the sample is subject to selection effects. Self-report measures may exhibit recall and social desirability biases; we mitigated these through low-burden design and anonymity. Platform-reported promotion metrics are not independently audited. Creator non-response limited triangulation with insider perspectives; we analyze this pattern as boundary-work rather than as missing data. Finally, the study is cross-sectional and anchored to August 8, 2024 through September 4, 2024, which may not capture dynamics outside that event window.

\section{Results}
\subsection{Survey Results}

We begin by orienting the reader to the dataset. Table~\ref{tab:themesummary} summarizes the themes, their representative codes, and coverage levels (general, typical, variant). For transparency, Appendix~\ref{tab:codetotheme} provides the complete code-to-theme mapping, and Appendix~\ref{tab:jointdisplay} presents a joint display that aligns quantitative patterns, qualitative themes, and illustrative quotes. Together, these materials offer a concise roadmap. The subsections that follow develop each theme in depth and integrate survey distributions with selected quotes.

\begin{table*}[t]
\centering
\caption{Themes, example codes, and participant coverage. For the full code-to-theme mapping with counts, see Appendix~\ref{tab:codetotheme}.}
\setlength{\tabcolsep}{1pt}
\renewcommand{\arraystretch}{1.2}
\begin{tabular*}{\textwidth}{@{\extracolsep{\fill}} p{0.22\textwidth} p{0.56\textwidth} p{0.12\textwidth}@{}}
\toprule
\textbf{Theme} & \textbf{Example Codes} & \textbf{Coverage} \\
\midrule
Entertainment as Baseline & Trend/Entertainment; Stable/Repetitive Content; Shifts in Style & \hfill 43.6\% \\
Atmosphere as Contested & Supportive/Positive; Toxic/Negative; Mixed/Contested & \hfill 81.0\% \\
Belonging \& Outsider Positioning & Non-membership; Engagement-based Belonging; Offline/Online Belonging & \hfill 34.4\% \\
Parasocial Anchors & Identity Anchors (parasociality); Creator Centered & \hfill 13.0\% \\
Event Temporality & Continuation/Hope; Cyclical/Event-Based; Decline/Fading & \hfill 41.7\% \\
Offline Linkage \& Structural Critique & Impact on PNMs; Critiques of Greek System; Sponsorship/Costs & \hfill 11--14\% \\
\bottomrule
\end{tabular*}
\label{tab:themesummary}
\end{table*}

\subsubsection{Atmosphere as Contested and Ambivalent}
Respondents most often appraised RushTok through its atmosphere. In our codebook, this was the broadest theme (general coverage, 81\%), yet it was rarely described as a single mood. Participants depicted a space that could feel celebratory and supportive, yet also defensive, judgmental, and prone to pile-ons. They repeatedly identified tension in how spectators, creators, and the sorority system were policed in the comments and in meta-commentary about who belongs.

Debates about the sorority system set the tone for many. One participant captured the polarization bluntly: {``People defend Greek life to their deaths, which is how you know it’s toxic and needs to be abolished''} (P44). Others described how efforts to call out harassment invited retaliation: {``I commented that it was wrong to leave negative and toxic comments \ldots{} and was called creepy and parasocial''} (P58). These accounts highlight boundary-work in action, including defense of the institution or its critics and policing the legitimacy of those who speak up.

Spectatorship further complicated the tone. Several respondents reported being unsettled by the presence and intensity of older viewers, while acknowledging their own complicity as watchers. As one put it, {``It freaks me out a little, tbh \ldots{} it's often older women serving as the critics/commentators \ldots{} Maybe I'm not much better as a passive viewer''} (P63). Another noted an escalation from online attention to offline presence. {``There are way too many older women inserting themselves or getting overly invested \ldots{} it's very unsettling \ldots{} get in your car and drive to campus for bid day without having any reason to be there''} (P40). Together, these reflections cast the atmosphere as a moral terrain in which both who is watching and how they watch become objects of contention.

Moments of controversy amplified this ambivalence. Disagreements over high-profile PNMs, especially `Bama Morgan', foreground both solidarity and hostility. One respondent observed rallying and attack in the same breath: {``they are mostly rallying behind @bamamorgan and attacking the other girls who are already in the sororities \ldots{}''} (P10). Another described a community split over why Morgan was {dropped} (\textit{i.e.}, rejected by all sororities), responding {``some people thought it made sense due to her large social media presence, others thought she did not receive a bid due to `looking different'''} (P53). These dynamics made debate itself a feature of the atmosphere, not a deviation from it.

Nostalgia and disappointment surfaced as counterpoints to ongoing conflict. Some longed for a lighter register, asking {``can't we just get back to pretty dresses and celebrating sisterhood.''} (P42). Others described a shift toward more surface-level content, experienced as a flattening of tone: {``Less women were willing to tell us about their experiences. Most of the content was pretty superficial this year''} (P18). These sentiments do not erase contention; they sit alongside it, marking a space that oscillates between spectacle, critique, and a desire for civility.

Overall, participants treated atmosphere as the primary lens for understanding RushTok: contested, negotiated, and frequently ambivalent. Defenses of the institution, accusations of overreach, discomfort with spectatorship, and recurring flare-ups around specific figures coexisted within the same event

\subsubsection{Entertainment as the Baseline of Engagement}
Survey responses frame RushTok primarily as feed-delivered entertainment. Most encountered it through the \textit{For You} page (54/71), with smaller groups arriving by search (13), other social media (3), or a friend (1). Viewed content clustered in formats optimized for rapid, serial viewing, including outfit-of-the-day videos (61/71), rush-week experiences (48/71), and dance content related to rush (47/71). These distributions align with how viewers described the experience---as something to watch, sample, and revisit. One participant called it {``fascinating to watch for the month of August''} (P65), while another described themselves as {``just an observer \ldots{} it's fun/interesting to see a different lifestyle''} (P24).

Entertainment here is not mere diversion; it is a register shaped by polish, repetition, and spectacle. Respondents noted a production shift, {``seemed more produced this year''} (P5), alongside a sense of iterative sameness, {``it feels repetitive, but people still check in every August''} (P33). Financial spectacle also kept people watching: {``I was more so engaged because I’m just in shock about the amount of money involved''} (P27). Together, these reflections describe an experience that is easy to consume and hard to ignore: polished yet formulaic, familiar yet attention-holding.

Behavioral indicators point to light-to-moderate engagement within this entertainment frame. A majority interacted with at least some videos through likes, comments, or shares (60/71 selected {A few}, {Some}, {Most}, or {All}), yet following remained limited. Thirty-one reported following none of the creators they encountered, thirty-one followed only a few, and just nine followed some, most, or all. This pattern suggests a feed-attuned consumption mode: frequent touchpoints, modest commitment, and creator affinity that rarely converts to deeper, more durable ties.

Taken together, the survey points to entertainment as the default mode of encountering RushTok. Algorithmic delivery foregrounds short, repeatable formats; production value and financial spectacle hold attention; interaction is frequent yet light. From this baseline, claims of belonging and creator attachment make sense: many remain spectators, and a smaller subset engages more deeply.

\subsubsection{Belonging and Outsider Positioning}

Recognition of collective life outpaced personal affiliation: 51/71 respondents agreed RushTok could be described as a community, yet only 11/71 identified as members. This gap also appeared in open responses, where participants drew boundaries around what counts as belonging and how it should feel.

Several respondents framed community as reciprocity rather than reach. As one put it, {`` I don’t consider RushTok a community. Communities involve bidirectional conversation. The posters rarely, or never, converse with the viewers''} (P54). Others limited membership to those inside the ritual itself, {``I consider rushtok the community of girls actively in a sorority and rushing for one as well''} (P13). A third position cast viewers as adjacent but not included, listing community members as {``PNMs, Actives, Alumni, `Rush adjacent' (parents, coaches), etc.''} and identifying as {``just an observer. I don't post, like, or comment. Just watch.''} (P40).

Thresholds for belonging were often described as conditional or achieved. Some asked whether passive watching was enough, {``Am I part of a community for just watching videos?''} (P5). Others pointed to engagement with specific PNMs as the tipping point, {``when I started interacting more with the PNMs and began paying more attention to specific PNMs, hoping to see where they would run home to!''} (P55). For some, belonging required personal resonance beyond the feed, {``when it includes an aspect of my personal life/a community I would connect with IRL''} (P24).

Belonging also moved with the event’s tone. One participant contrasted last year's feeling with this year's divisions, declaring {``it’s far too divided and negative this year to be a community. If you had asked me last year, I may have said yes. I feel discouraged this year''} (P58). Others defined community through support norms and judged RushTok as falling short, {``a community to me offers support to others. While I saw some of that, it all felt very fake to me. I consider [RushTok] a group of people that are overly critical of teenagers on the internet''} (P27).

Taken together, these patterns show layered belonging. Many acknowledged a shared public space, but claims of membership hinged on reciprocity, contribution, proximity to the offline ritual, or meaningful identification with particular people and stories. Outsider positioning remained common---watching without joining---was common and dynamic; it shifted with perceived civility, with the salience of specific PNMs, and with whether participation felt personally relevant or reciprocated.

\subsubsection{Parasocial Anchors and Creator Attachment}
If RushTok coheres, it does so because specific creators become its main characters. Viewers did not gather around an abstract rush culture so much as around a few recognizable PNMs whose personalities and perspectives turned recruitment into a seasonal storyline. These asymmetrical ties organize attention and action. They decide who is watched, who is defended in moments of controversy, and who carries audience loyalty from one season to the next. In our coding, parasocial anchors captured this creator-centered gravity (variant coverage, 13\%).

In the multiple-choice items, respondents most often said they liked creators for personality (52/71) and for perspective on the rush process (37/71). Only one described affinity with `rush culture' itself, indicating that identification tended to run through individuals who became recognizable protagonists.

Free-text responses make these anchors visible. Some credited singular figures with catalyzing attention, such as one participant declaring {``Kylan Darnell remains the reason Alabama ZTA got put on the map''} (P12). Others mapped the audience by the kinds of parasocial ties that formed, describing RushTok as {``two separate communities---PNMs and college students, and older adults who are living vicariously through them''} (P50). This split helps explain why creator-centered interest persisted even when the broader phenomenon felt familiar or repetitive.

Creator attachment also shaped how controversy landed on the participants and PNMs. One respondent read institutional responses as a reaction to outside parasocial intensity, saying {``The Bama sorority side of rushtok is very sensitive to overly negative outside feedback and is defensive (e.g., dropping PNMs with large parasocial followings from outside older adults)''} (P50). Another connected the current audience to those lingering attachments, not to renewed curiosity about rush, describing RushTok as {``somewhat stale. Outside older adults are no longer `curious'---they know what rush at bama looks like. At this point mostly those with parasocial attachments to these PNMs are left, and those who have been successful with `commentary' on the cost of the OOTDs''} (P50).

Put plainly, creators function as the main characters that organize RushTok. Personalities and perspectives draw viewers in; parasocial ties decide who keeps watching, who steps in during controversy, and who earns loyalty across seasons; institutional reactions follow when attention concentrates on particular PNMs. Treating RushTok as cast-driven helps explain why interest persists even when the broader ritual feels familiar: the audience comes back for returning leads and new arcs, while the algorithm spotlights which storylines rise or recede.

\subsubsection{Ephemerality, Event Temporality, and Seasonal Rhythm}

Respondents described RushTok as a cyclical surge that returns with recruitment and recedes in short order. Exposure spanned multiple seasons for most participants, with 52/71 reporting watching across more than one year. Among those multi-year viewers, 38/52 said they saw the same creators return year to year and 14/52 did not. Day-to-day variety still characterized many feeds (44/71 reported seeing a ``variety of content creators'', while 27/71 saw ``mostly the same''), which helps explain how the event can feel both familiar and new in each cycle.

Participants also traced stylistic drift across seasons. One respondent contrasted early, rough-cut videos with later polish, noting that {``2021 videos were unpolished and relatable. Each year they became more and more posed''} (P54). Others described a shift in topical mix toward display over process: {``This year was mainly OOTD. Used to see more insight on the process, experiences, weather, etc''} (P31). For some, the cycle now required effort to locate, with one participant reporting that they {``Had to search for it more this year \ldots{} like the excitement is waning/getting redundant''} (P42). These observations point to a short attention half-life and growing discoverability friction as the surge fades.

Taken together, the patterns depict a seasonal but ephemeral rhythm: attention peaks quickly during rush, then decays as the feed drifts. Multi-year viewers recognize returning leads, the feed still supplies variety, yet rising production polish and a narrowing of formats make each revival feel more predictable. Ephemerality is therefore not absence but cadence, that is, brief, high-intensity windows that recur annually, sustain interest through returning characters, and then recede until the next cycle.

\subsubsection{Limited Offline Linkage and Structural Commentary}
Participants largely encountered RushTok on their phones rather than in offline settings. Most had no formal tie to the University of Alabama or Greek life (63/71 reported being ``Not affiliated''), and most said they experienced the community only online (56/71 ``Only online/on TikTok'' vs. 15/71 ``In real-life as well''). These distributions frame how respondents reflected on institutions, visibility, and control.

Even without direct affiliation, respondents offered structural readings of how RushTok is produced. Several read the trend through costs and participation, noting increased costs and fewer posters. As one put it, they saw {``more money \ldots{} being spent''} and {``less creators posting than in the past''} (P20). Others pointed to rules and policy-driven opacity, interpreting guardedness as deliberate rather than disinterest: {``it’s \ldots{} getting more vague about details \ldots{} Panhellenic \ldots{} want to keep the rush process `secret' so each PNM can have their own experience''} (P8). Posting style shifted accordingly. One participant observed that {``[2024 PNMs] seems more reserved/discreet in the content they're sharing''} (P40).

Structural commentary also touched on who gets seen and amplified within RushTok. Some respondents described a broader range of PNMs appearing, reading it as a gradual diversification of the on-screen cohort. One participant noted that {``there's more diversity in the girls who are posting and i am seeing more than just bama rush''} (P10). Others emphasized the event’s annual cadence as the dominant offline linkage, with a new academic year bringing new creators: {``It's the start of a new academic year, new ladies starting up their posting. And we're here for it and here to meet them!''} (P56).

Taken together, these accounts depict a public that remains primarily online for most viewers yet is shaped by offline governance and resource dynamics. Limited personal affiliation and online-only encounters coincide with perceptions of tighter institutional control, selective sharing by PNMs, and gradual shifts in who appears on camera. The result is a mediated view of a tightly managed ritual: visible at scale in the feed, thinly connected to everyday campus life for most respondents.

\paragraph{Synthesis.}
Across themes, respondents describe RushTok as an algorithmically assembled public met primarily via the \textit{For You} page (54/71) and revisited each August (52/71). The baseline is entertainment with a mixed tone. Affiliation centers on a small set of recognizable PNMs: viewers like creators for personality (52/71) and perspective (37/71), while affinity with so-called rush culture is rare. Many recognize a collective but stop short of joining it; a majority call it a community (51/71) yet few claim membership (11/71). Interaction is common but light (60/71 engage at least a little) and seldom converts to following (31/71 follow none). The cadence sustains attention but reduces novelty, with returning leads common among multi-year viewers (38/52) and some reporting waning excitement. Offline ties are thin, consistent with limited UA affiliation (63/71) and online-only experience (56/71). Taken together, RushTok functions less as a durable community than as a recurring spectacle anchored by creators and the platform’s feed.

\subsection{Interview Attempts: Silence and Insularity}
Despite multi-channel outreach, fewer than 10\% of invitations received a response, most replies declined comment, and no creators consented to be interviewed. Some PNMs explicitly cited concern that participation could affect recruitment outcomes.

We treat this silence as data. It sits in tension with the survey portrait of high visibility and easy audience access via the \textit{For You} page. Viewers can encounter and discuss RushTok readily, yet those producing content maintain strict boundaries against external engagement. This asymmetry suggests creator-led boundary-work that keeps the public open to spectators but closed to outsiders.

We report this non-response as a substantive result rather than a procedural limitation. It indicates layered community boundaries and a governance logic in which visibility is curated and contact is controlled. We return to these implications in the Discussion, where we consider what such gatekeeping means for participant protection, platform design, and research access.

\section{Discussion}
\label{sec:Discussion}
RushTok represents an archetype of what we call event-based algorithmic communities:  algorithmically assembled collectives (cf.\ \cite{Papacharissi-AffectivePublicsSentiment-2014}) that form around bounded offline rituals such as recruitment week \cite{Turner-RitualProcessStructure-2011}. Their persistence rests on episodic attention, ambivalent affect, and seasonal recurrence rather than durable social ties. This model contrasts with network-based communities, which are sustained by ongoing relationships, and identity-based communities, which hinge on stable affiliation and belonging (cf.\ \cite{Wenger-CommunitiesPracticeLearning-1999}).

\subsection*{Finding F1: Ambivalence as circulation.}

Ambivalence is not a side note in RushTok. It helps keep the community in motion. By ambivalence, we mean the co-presence of positive and negative framings about the same thing.

Participants did not report a singular mood. They described warmth, humor, and shared enjoyment alongside exclusion, toxicity, and critique, often within the same response. A respondent might laugh with a creator and also recoil at conspicuous privilege, or defend a favorite while noting that the broader tone can be {mean}. This pattern was central, not rare, and it was a primary way people made sense of RushTok.

Ambivalence also fits how the feed invites interaction. Praise draws sympathizers and playful participation. Critique draws rebuttals, meta-arguments, stitches, and duets. Both routes generate comments and repeat exposures that can increase distribution. We do not claim knowledge of ranking internals; instead we propose a routine pathway consistent with our data and with visible features of TikTok: divergent reactions can lead to wider reach. In this sense, ambivalence works as a resource for visibility and return visits.

We argue that this dynamic helps to explain the membership gap revealed in the survey. Here we distinguish two senses of ambivalence. \textit{Affective ambivalence} refers to mixed feelings about the same content (praise and critique co-present). \textit{Belonging ambivalence} refers to recognizing a collective while withholding membership. In our data, 51 of 71 respondents agreed that RushTok could be described as a community, yet only 11 of 71 identified as members. The same features that generate collective energy (inside jokes, shared favorites, rapid circulation) also introduce volatility that discourages commitment, producing belonging ambivalence. Many participants explicitly positioned themselves as outsiders, asking {``Am I part of a community for just watching videos?''} (P5). This stance was not passive. By watching, liking, and sometimes commenting, these non-members sustained visibility. Outsider positioning was therefore generative: it produced circulation while resisting self-identification.

We describe this pattern as \textit{layered belonging}. Participation is graduated rather than binary, ranging from casual viewers, to commenters fluent in insider humor, to a small set of self-identified members. Each layer contributes to circulation, including those who disavow membership. This unsettles a common view in that treats atmosphere as a background condition that either enables or inhibits engagement. In RushTok, movement between support and critique is part of participation. Our claim is intentionally bounded: in event-based algorithmic communities, ambivalence operates as an ordinary route to attention rather than an anomaly.

A note on instrument effects. Because the survey asked about perceived change and the current state of the community, respondents oriented to tone and atmosphere at the collective level. We treat this as analytically useful. When asked to make sense of RushTok, participants reached first for atmosphere rather than for creator rosters or offline consequences. That emphasis links the two forms of ambivalence defined above (affective and belonging) and motivates the analyses that follow on belonging, parasocial anchoring, and seasonal persistence.

\subsection*{Finding F2: Seriality as persistence.}

In event-based algorithmic communities, seasonal recurrence of the offline ritual and platform-led serialization of content together substitute for durable membership as the mechanism of persistence.

If ambivalence shaped the day-to-day experience of RushTok, spectacle plus recurrence and serialization shaped its overall form. Participants described the feed as repetitive yet addictive, polished yet formulaic (``Seemed more produced this year'' (P5) but also ``it feels repetitive, but people still check in every August'' (P33)). This mix shows how TikTok packages an offline ritual into a serialized storyline that returns each year. Sorority recruitment week becomes not only a campus practice but an annual season that viewers can drop into and follow without joining a group.

Entertainment was not incidental; it organized attention and drew in viewers with no ties to sorority life. What Turner describes as ritual process \cite{Turner-RitualProcessStructure-2011} (heightened visibility, shared performance, climactic resolution) was recast through algorithmic curation as short, repeatable clips. The result is not only coverage of an event but a storyline that can be followed; spectatorship itself becomes a mode of participation.

Empirically, the feed presents recurring roles and cues. PNMs appear as leads; OOTDs function as recurring props; bid day operates as a predictable climax that resets the arc. Multi-year viewing was common (52/71), and among those viewers many reported seeing the same creators return (38/52). Viewers can drop in, recognize the pattern, and continue without joining a group. We conclude that persistence comes from recurrence and serialization rather than member ties. In Turner's terms, RushTok takes the form of a seasonal ritual, a bounded liminal interval whose intensity is concentrated in time \cite{Turner-RitualProcessStructure-2011}.

RushTok persisted through its recognizable storyline and annual recurrence. Its collectivity was not built from strong bonds among members but from the ritual return of an entertaining spectacle. For platform design and governance, this raises pressing questions: What responsibilities do platforms have when their algorithms transform offline rituals into serialized entertainment? How should they scaffold participation for those thrust into the spotlight, even as they optimize for spectacle to draw in audiences? We point to two concrete patterns that we elaborate on in Implications: (1) graduated visibility controls and (2) context banners for sensitive rituals.

\subsection*{Finding F3: Amplification concentrates risk.} 
Algorithmic amplification elevates a few participants into sudden prominence. Attention clusters around identifiable PNMs, which concentrates vulnerability on them while distributing entertainment to many others.

Survey responses point to creator-centered affiliation rather than attachment to ``rush culture.'' Most respondents valued creators for personality (52/71) and for perspective on the process (37/71), with almost no one citing affinity with the institution itself. Free-text accounts named specific figures and described audience mobilization around them. One respondent credited a single breakout: ``Kylan Darnell remains the reason Alabama ZTA got put on the map'' (P12). Another observed coordinated support and antagonism during controversy: ``they are mostly rallying behind @bamamorgan and attacking the other girls who are already in the sororities'' (P10). A third mapped the audience by the kinds of ties that formed, noting ``Two separate communities—PNMs and college students, and older adults who are living vicariously through them'' (P50).

Concentrated attention also drew institutional caution and public scrutiny. Respondents perceived defensiveness toward highly visible PNMs, including reports that chapters reacted to large parasocial followings (P50). Others recalled disagreement over why a well-known PNM was dropped, with explanations ranging from her social media presence to her appearance (P53). Additional comments pointed to more discreet posting by the current class (P40) and guidance to keep process details off-platform (P8). Together, these observations explain why bigger online footprints can threaten social standing offline. Visibility invites judgment from audiences and risk management by organizations.

For most viewers, stakes were low. Many had no affiliation with Alabama or Greek life (63/71) and experienced RushTok only online (56/71). They watched, sometimes interacted, and moved on. For those at the center, exposure persisted after the season ended and was tied to reputational futures. This asymmetry is a defining feature of event-based algorithmic publics. Platforms distribute collective attention widely, yet the burdens of amplification are borne by a small set of visible individuals.

This reading clarifies what community means under creator anchors. Affect gathers around a few recognizable people who become both emblem and target. The collective remains large, but its gravity is localized. Design and governance should therefore recognize and support participants who are pushed into instant visibility. Practical steps include graduated visibility during surge windows, prompts and tools for creator safety, contextual banners that mark sensitive time-bounded rituals, and aftercare resources once the event recedes.

\subsection*{Finding F4: Protective ritual scope condition.}

When the offline ritual is guarded, event-based publics look open to viewers yet feel defensive to those on camera. Claims from such cases should not be generalized to celebratory or activist events without comparative study.

RushTok unfolded within the institutional culture of Greek life, which is marked by secrecy and tight control. Respondents described a turn toward restraint and opacity: ``[2024 PNMs] seems more reserved/discreet in the content they're sharing’’ (P40). Others linked caution to guidance from organizers, noting that panhellenic bodies encourage keeping process details off platform so PNMs can have their own experience (P8). Participants also perceived a shift toward surface-level posts and fewer first-person accounts, with one participant noting that ``this year was mainly OOTD. Used to see more insight on the process’’ (P31).

These patterns sit alongside limited personal ties to the institution. Most respondents reported no affiliation with Alabama or Greek life (63/71) and experienced the community only online (56/71). Viewers could watch freely, yet creators signaled boundaries and shared less. Our auto-ethnographic attempt to interview creators reinforces this reading. Despite repeated outreach through public contacts and snowballing, no one agreed to participate. We treat this ``productive failure’’ as evidence of insulation rather than as a recruitment error.

Placing these observations together clarifies earlier findings. Ambivalence, layered belonging, and thin offline linkage are not free-floating traits. They reflect a protective ritual under sudden visibility. The algorithm opens the curtain and participants at the center respond with boundary-work that manages risk and reputation.

This scope condition matters for theory and design. Not all event-based publics will display the same defensive posture. Concerts, sports tournaments, and political rallies often invite attention and contact. Protective, celebratory, and contentious rituals likely produce different dynamics once they meet ranking-driven feeds. Comparative studies across these settings are needed to characterize that range.

Design and governance should also adapt to protective contexts. Graduated visibility during surge windows, context cues that mark sensitive rituals, and low-friction research and reporting channels can reduce pressure on individuals while the event is most visible. Our recommendations are tailored to cases where institutional secrecy and reputational stakes shape what participants are willing to show.

\subsection*{Finding F5: Silence as boundary signal.}

In algorithmically assembled publics, non-response is interpretable boundary-work; treating it as data (and not deficit) reveals how communities govern exposure and resist investigation.

\begin{table*}[t]
\centering
\caption{Governance blueprint for event-based algorithmic communities. Mechanisms (from Section \ref{sec:Discussion}) are linked to risks, platform levers, and operational triggers/decays suitable for surge-style events like RushTok.}
\label{tab:governance-blueprint}
\small
\setlength{\tabcolsep}{4pt}
\renewcommand{\arraystretch}{1.15}
\begin{tabular}{p{0.13\textwidth} p{0.20\textwidth} p{0.35\textwidth} p{0.24\textwidth}}
\toprule
\textbf{Mechanism} & \textbf{Risk} & \textbf{Platform lever (design/policy)} & \textbf{Trigger / Decay} \\
\midrule

\textbf{P1: Ambivalence as circulation} &
Negativity fuels engagement spikes; pile-ons, brigading, dogpiling in comments. &
\emph{Graduated visibility} during surges: rate-limit comments/replies; friction for stitches/duets; civility nudges; temporary quorum rules (e.g., first-time commenter cooldown); interstitials on highly polarizing threads. &
\textbf{Trigger:} rapid growth in views/comments for event tags (e.g., \#RushTok) above baseline over 24\,h; \textbf{Decay:} auto-lift after $>$48--72\,h below threshold; manual override for edge cases. \\[0.6ex]

\textbf{P2: Seriality as persistence} &
Out-of-context resurfacing post-event; outdated claims continue to propagate; context collapse. &
\emph{Context banners} (``Rush Week 2025''); time-scoped recommendation decay (half-life after event end); de-amplify stitches of archived clips; optional creator prompt to archive or disable remix after event. &
\textbf{Trigger:} event window start (calendar or burst detection); \textbf{Decay:} scheduled recommendation half-life and banner timeout (e.g., 14--30 days). \\[0.6ex]

\textbf{P3: Parasocial concentration \& uneven risk} &
Sudden spotlight on a few individuals; harassment/doxxing; inbox flooding; reputational harm. &
\emph{Aftercare toolkit}: one-click privacy review; default stronger comment filters; follower-only comments; auto-block phrase lists seeded by event lexicon; optional ``slow mode'' for DM requests; creator-facing safety analytics. &
\textbf{Trigger:} creator-level velocity thresholds (e.g., impressions/day above baseline; follow spikes); \textbf{Decay:} auto-revert when velocity normalizes for 7 days; user-controlled persistence. \\[0.6ex]

\textbf{P4: Protective ritual / offline linkage} &
Misinterpretation of insider practices; external surveillance of physical spaces; location-based risks. &
\emph{Ritual sensitivity affordances}: educational info panel on the event; default blur/suppress precise locations; friction on reposting private-space footage; ``share with context'' prompts on re-uploads. &
\textbf{Trigger:} detection of event tags plus near-venue geo signals; \textbf{Decay:} ends at event close (e.g., Bid Day) plus short tail (7--14 days). \\[0.6ex]

\textbf{P5: Silence as boundary-work (research access)} &
Extractive outreach; misrepresentation; creator burden from unsolicited requests. &
\emph{Mediated research channel}: opt-in creator registry for outreach; templated ethics attestation; capped contact frequency; aggregated ``broadcast call'' instead of DMs; one-click ``decline and mute research requests''. &
\textbf{Trigger:} platform-whitelisted study or verified academic request; \textbf{Decay:} auto-expire per study window or event end; creator can persistently opt-out. \\[0.6ex]

\textbf{Spectacle-driven discoverability} &
Rumors/misinformation accelerate; low-signal amplification. &
\emph{Event-specific circuit breakers}: claim-type interstitials; accelerated fact-check routing; temporary share-limit on flagged claims; link-out to authoritative explainer. &
\textbf{Trigger:} cluster of reports/flags or anomaly in claim re-uploads; \textbf{Decay:} lift when claim volume and flags subside for 48\,h. \\[0.6ex]

\textbf{Creator and bystander well-being} &
Burnout and regret post-surge; lack of recovery tools; bystander fatigue. &
\emph{Aftercare and recovery}: post-event dashboards summarizing reach; nudges to review past posts; batch disable of stitches/duets; well-being resource surfacing; optional auto-archive. &
\textbf{Trigger:} event end plus creator velocity drop; \textbf{Decay:} sunset after 14--30 days or upon creator dismissal. \\
\bottomrule
\end{tabular}
\normalsize
\end{table*}

As noted in the Results, our attempts to recruit RushTok creators for interviews yielded only silence or polite refusals. We interpret this outcome not as missing data but as productive failure \cite{Kapur-ProductiveFailure-2008}: evidence of how algorithmically assembled publics actively govern their boundaries.

Creators’ non-response functions as boundary-work \cite{StarGriesemer-InstitutionalEcologyTranslations-1989,BowkerStar-SortingThingsOut-}. Whereas casual viewers could access RushTok effortlessly via the For You Page and even self-identify as members, creators maintained strict limits on external engagement. Silence was not absence but agency—an assertion of control in the face of involuntary visibility. For participants thrust into instant influencer roles, refusal operated as risk management against scrutiny, misrepresentation, and reputational harm.

These refusals also foreground ethical stakes for HCI and CSCW research. Publicly available contact information does not equate to consent; silence may indicate vulnerability or reputational caution. Treating non-response as meaningful and refraining from persistent follow-up is an ethical stance aligned with calls for sensitivity when studying precariously visible populations \cite{Markham-FABRICATIONETHICALPRACTICE-2012,FieslerProferes-ParticipantPerceptionsTwitter-2018}.

Methodologically, RushTok underscores the limits of conventional qualitative access in algorithmic contexts. Interviews may fail not because recruitment is poorly designed, but because insulation is constitutive of these publics. Documenting such failures thus contributes to methodological reflexivity: non-response can be data about governance and boundary maintenance in algorithmic communities.

\subsection{Implications for HCI and Platform Governance}

To make our implications deployable, we present a governance blueprint that maps mechanisms of event-based algorithmic communities to concrete risks, levers, and operational triggers/decays (Table~\ref{tab:governance-blueprint}).

\textit{How to read the blueprint.} For P1 (ambivalence as circulation), the lever bundle is graduated visibility during surges, triggered by anomalous tag-level velocity and lifted after activity decays for 48–72h. For P3 (parasocial concentration), the aftercare toolkit is creator-scoped and reverts when creator-level velocity normalizes or at user discretion. Across rows we emphasize reversible, time-bounded interventions, creator control where feasible, and transparency (context banners).

Taken together, these interventions recognize ambivalence as an organizing logic, scaffold sudden visibility, mitigate uneven stakes, and govern ephemeral, event-based publics as first-class governance targets. In practice, they translate our analysis into concrete, time-bounded levers—activated by clear surge triggers and retired on decay—while preserving creator control and attending to participant vulnerability.

\subsection{Limitations \& Scope Conditions}

Our analysis is interpretive and descriptive. We used reflexive thematic analysis with coverage labels to mark patterning in the corpus. These labels do not imply prevalence in a wider population. Quantitative summaries are descriptive and small-n contrasts, such as the self-identified member group, are unstable. We did not estimate causal effects.

Instrument design also matters. The survey asked about change over time and about the current state of the community. Those prompts likely foregrounded atmosphere, seriality, and creator focus. We defined membership through self-identification and exposure through self-estimates. These operational choices privileged sense-making at the collective level rather than granular behavior.

Several data sources were out of scope. We did not analyze behavioral logs, content corpora, comment networks, or ranking internals. Claims about circulation mechanisms are therefore theorized from observed affordances and respondent accounts, not from audits of the recommendation system. We also did not link individual survey responses to specific posts or creators, which limits event-level triangulation.

These boundaries do not diminish the contribution; they specify where the theory holds and invite comparative tests across different event types and institutional cultures.

\section{Contributions}
This paper makes the following contributions:
\begin{enumerate}
    \item \textbf{Conceptual:} We define \emph{event-based algorithmic communities} as a distinct form of collective sociality---\emph{temporally bounded} publics sustained by ambivalence, spectacle/seriality, and algorithmic amplification rather than durable membership.
    \item \textbf{Empirical:} We offer an empirical account of RushTok, showing how layered belonging, parasocial anchors, and \emph{protective boundary-work} shape participation and visibility.
    \item \textbf{Methodological:} We advance \emph{productive failure} as a methodological lens, demonstrating how non-response and silence can evidence community insulation and the limits of conventional qualitative access.
    \item \textbf{Design \& Governance:} We identify responsibilities for platforms in scaffolding ephemeral publics, including designing for ambivalence, supporting individuals thrust into sudden visibility, addressing uneven stakes of amplification, and governing publics that are short-lived yet high-stakes.
\end{enumerate}

\section{Conclusion}
Placing RushTok in view clarifies how platforms gather people around rituals and how those gatherings feel from the feed and the ground. The paper names this form---\emph{event-based algorithmic communities}---and traces how ambivalence, spectacle, and timing hold attention without demanding durable membership. The account is anchored in evidence and written for reuse: it connects familiar theories of ritual, fandom, and practice to contemporary distribution logics, and it offers language that travels beyond a single event or platform.

The contributions outlined come together as a path rather than a checklist. Conceptual framing sets the stage for empirical description, which in turn motivates a reflexive method when access closes down. Design implications then follow in the same spirit: pace amplification during peaks, add context when a ritual is sensitive, support people who did not plan to become visible, and help them manage traces once the event recedes. None of these steps asks platforms to dull what makes these publics compelling. Instead they ask for care where attention concentrates and for tools that match the tempo of short-lived events.

A better trajectory is within reach. Platforms can recognize event temporality, foreground consent and context, and tune visibility when attention surges. Researchers can build comparative and longitudinal programs that follow publics across a variety of events, pairing surveys, reflexive accounts, audits, and in-situ traces. By naming the form, demonstrating its dynamics, and outlining practical levers, this paper opens a path for studying and stewarding algorithmically assembled publics that are short-lived yet consequential. The promise is a social web where visibility still delights and surprises, and where those at the center are better supported when the spotlight turns their way.

\bibliographystyle{ACM-Reference-Format}
\bibliography{biblio}

\appendix
\onecolumn
\section{Appendix Tables}

\begin{table}[h]
\centering
\caption{Inter-coder agreement by coder pair and code type. Values are pairwise percent agreement. 
\emph{Base} refers to base codes; \emph{Additive} refers to secondary codes applied in addition to base codes; 
\emph{Overall} is unit-level agreement across all codes after enforcing base/additive dependencies. 
We report this descriptively to foreground transparency; we do not treat agreement as a validity criterion.}
\label{tab:icr}
\begin{tabular}{lccc}
\toprule
\textbf{Coder pair} & \textbf{Base} & \textbf{Additive} & \textbf{Overall} \\
\midrule
A--B  & 84.2\% & 94.7\% & 79.0\% \\
A--C  & 75.6\% & 97.6\% & 73.2\% \\
B--C  & 86.5\% & 100.0\% & 86.5\% \\
\midrule
\textbf{Mean across pairs} & 82.1\% & 97.4\% & 79.5\% \\
\bottomrule
\end{tabular}
\end{table}


\begin{table}[]
\centering
\caption{Full mapping of codes to higher-level themes with counts from the consensus-coded dataset. 
This appendix table complements the summary view in Table~\ref{tab:themesummary} by providing detailed transparency into how codes informed theme construction.}
\begin{tabular}{llr}
\toprule
\textbf{Theme} & \textbf{Code} & \textbf{Count} \\
\midrule
\textbf{Entertainment as Baseline of Engagement} & Trend/Entertainment & 20 \\
 & Stable/Repetitive Content & 23 \\
 & Shifts in production and style & 18 \\
 & OOTD/Fashion Emphasis & 8 \\
 & High spending/Luxury emphasis & 5 \\
 & Reduction in sharing/Secrecy & 7 \\
 & Diversity/inclusion emphasis & 3 \\
 & Sponsorship/Branding & 2 \\
\midrule
\textbf{Atmosphere as Contested and Ambivalent} & Supportive/Positive & 37 \\
 & Toxic/Negative & 40 \\
 & Mixed/Contested & 55 \\
 & Cliquey/Exclusive & 8 \\
 & Nostalgia & 9 \\
\midrule
\textbf{Belonging and Outsider Positioning} & Non-membership/outsider positioning & 15 \\
 & Engagement-based belonging & 14 \\
 & Offline/Online Belonging & 10 \\
\midrule
\textbf{Parasocial Anchors and Creator Attachment} & Identity Anchors (parasociality) & 13 \\
 & Creator Centered & 10 \\
\midrule
\textbf{Event Temporality and Seasonal Rhythm} & Continuation/Hope & 22 \\
 & Decline/Fading Interest & 18 \\
 & Cyclical/Event-Based & 17 \\
 & Fragmentation & 6 \\
 & Mixed/Contested Future & 5 \\
\midrule
\textbf{Limited Offline Linkage and Structural Commentary} & Impact on PNMs & 6 \\
 & Reputation of University/Greek Life & 4 \\
 & Offline/Online Disconnect & 2 \\
 & Critical of Greek system & 10 \\
 & Commodification/Sponsorship & 6 \\
 & Financial Costs/Barrier & 4 \\
 & Gendered Expectations & 3 \\
 & Racial Exclusion/Whiteness & 2 \\
\bottomrule
\end{tabular}
\label{tab:codetotheme}
\end{table}


\begin{table}[h]
\caption{Joint display of quantitative patterns, qualitative themes, and illustrative quotes from survey responses. Coverage is reported as general, typical, or variant to indicate prevalence across participants.}
\centering
\begin{tabular}{|p{3cm}|p{5cm}|p{5cm}|}
\hline
\textbf{Quantitative Pattern} & \textbf{Qualitative Theme (coverage)} & \textbf{Exemplar Quote} \\
\hline
Many respondents agreed RushTok is a community, but fewer said they were part of it & \textit{Spectatorship vs. Membership} (general) & ``I watched every video, but I wouldn’t say I was *in* RushTok. It felt more like watching a TV show than joining a group.'' \\
\hline
High exposure via For You Page; most engagement was passive (viewing, liking) & \textit{Entertainment / Curiosity} (typical) & ``It was fun and addictive — I scrolled for hours just to see what people wore, but I never commented or posted.'' \\
\hline
Smaller group self-identified as community members; these users engaged more deeply (following, commenting) & \textit{Belonging / Identity Anchoring} (typical) & ``We made jokes in the comments and it felt like everyone understood them. That made me feel part of something, even if I wasn’t rushing.'' \\
\hline
UA affiliation associated with perceiving RushTok as both online and offline & \textit{Event-Embeddedness / Impact on PNMs} (variant) & ``As someone on campus, you couldn’t escape it. The videos made the pressure even worse for the PNMs.'' \\
\hline
Viewers described being invested in specific PNMs despite no direct ties & \textit{Parasociality / Voyeurism} (variant) & ``I rooted for certain girls like they were characters in a show. I even felt nervous for them on bid day.'' \\
\hline
Critical reflections surfaced around Greek life and exclusivity & \textit{Critique of Greek System / Cliquey Atmosphere} (typical) & ``It all felt very elitist — like a spectacle of privilege — so I watched but never felt like I belonged.'' \\
\hline
\end{tabular}
\label{tab:jointdisplay}
\end{table}

\begin{table}[]
\caption{Demographics of survey participants}
\begin{tabular}{p{2cm}p{2cm}p{.2cm}p{2cm}p{.2cm}p{2cm}p{.2cm}p{2cm}p{.2cm}}
\toprule
{\textbf{Demographics}} \\ 
\bottomrule
{\textbf{University of Alabama Connection}}&{Not affiliated}&{63}&{Student}&{3}&{Family member of student}&{2}&{Alumni}&{1}\\
\midrule
{\textbf{Age Range}}&{18-22}&{20}&{23-34}&{21}&{35-44}&{19}&{45-54}&{8}\\
{}&{55-64}&{2}&{65+}&{1}\\
\midrule
{\textbf{Highest level of education}}&{High school/GED}&{6}&{Current Undergraduate}&{17}&{Associate Deg.}&{1}&{Bachelor Deg.}&{26}\\
&{Master Deg.}&{14}&{Professional Deg.}&{3}&{Doctorate}&{4}\\
\midrule
{\textbf{Gender}}&{Female}&{67}&{Male}&{3}&{N/A}&{1}\\ 
\bottomrule
\end{tabular}
\label{tab:demographics}
\end{table}

\end{document}